\documentclass{article}
\usepackage{cite}
\usepackage{amsmath,amssymb,amsfonts}
\usepackage{algorithmic}
\usepackage{graphicx}
\usepackage{textcomp}
\usepackage{xcolor}
\usepackage{hyperref}
\usepackage{subcaption}
\usepackage{fancyhdr}

\title{Blind Deblurring using GANs}
\author{Manoj Kumar Lenka\\Kalinga Institute of Industrial Technology, Bhubaneswar
\and Anurag Mittal, Anubha Pandey\\Indian Institute of Technology, Madras}
\date{July 8, 2019}
\begin{document}
	\maketitle	
	\thispagestyle{fancy}
	\section{Abstract}
		Deblurring is the task of restoring a blurred image to
		a sharp one, retrieving the information lost due to
		the blur. In blind deblurring we have no information regarding the blur kernel.
		
		As deblurring can be considered as an image to
		image translation task, deep learning based
		solutions, including the ones which use GAN
		(Generative Adversarial Network), have been
		proven effective for deblurring. Most of them have
		an encoder-decoder structure.
		
		Our objective is to try different GAN structures and
		improve its performance through various
		modifications to the existing structure for supervised
		deblurring.
		
		In supervised deblurring we have pairs of blurred
		and their corresponding shrap images, while in the
		unsupervised case we have a set of blurred and
		sharp images but their is no correspondence
		between them.
		
		Modifications to the structures is done to improve
		the global perception of the model. As blur is non-
		uniform in nature, for deblurring we require global
		information of the entire image, whereas convolution
		used in CNN is able to provide only local perception.
		
		Deep models can be used to improve global
		perception but due to large number of parameters it
		becomes difficult for it to converge and inference
		time increases, to solve this we propose the use of
		attention module (non-local block) which was
		previously used in language translation and other
		image to image translation tasks in deblurring.
		
		Use of residual connection also improves the
		performance of deblurring as features from the
		lower layers are added to the upper layers of the
		model. It has been found that classical losses like
		L1, L2, and perceptual loss also help in training of GANs when added
		together with adversarial loss. We also concatenate
		edge information of the image to observe its effects
		on deblurring. We also use feedback modules to
		retain long term dependencies.
		
		\textbf{Keywords:} image to image translation, encoder-decoder structure, global perception, CNN, attention module, perceptual loss, residual connections, edge information, feedback module.
	
	\section{Introduction}
		\subsection{Background and Rationale}
			Deblurring is an problem in Computer Vision and Image
			Processing. Blurring can be caused by object motion,
			camera blur or out-of-focus. The task of deblurring is to
			generate a sharp given a blurred image, it can be
			considered as a special case of image-to-image
			translation. GANs \cite{Goodfellow}\cite{Isola_2017_CVPR} have shown good performance for
			several image to image translation task like super
			resolution, image inpainting, etc including deblurring.
			Learning based methods used in deblurring can be
			broadly classified into two type, one in which we estimate
			the blur kernel \cite{Gong_2017_CVPR} \cite{Sun_2015_CVPR}, and the other in which we generate
			the sharp image in an end to end fashion \cite{Nah_2017_CVPR}\cite{Kupyn_2018_CVPR}\cite{Noroozi}\cite{Ramakrishnan}\cite{Tao_2018_CVPR}\cite{Zhang_2018_CVPR}. GANs have
			been used mostly to generate image in an end-to-end
			fashion.
			
			GANs consists of two parts a generator and a
			discriminator. The generator tries to map to the target
			image, whereas the discriminator tries to differentiate
			between the generator and the actual target images. The
			goal of generator is to fool the discriminator, so that it
			can’t differentiate between the generated and target
			image. In case of deblurring we condition the generator
			by giving it the blurred image as input, instead of some
			random noise, from which it tries to generate a sharp
			image in order to fool the discriminator.
			
		\subsection{Problem Statement}
			Given a blurred image B, our goal is to predict a sharp
			image S, such that
			\begin{equation}
				B = S * K + N 
			\end{equation}
			
			Where K is the blur kernel, N is the noise, and * denotes
			convolution.
			
		\subsection{Objective of Research}
			Our objective is to improve existing GAN structure in
			order to increase their global reception (i.e extract
			information from the entire image and the relation
			between the different parts of the image) as convolution
			used in CNNs provide local perception. As most blur in
			real world are non-uniform in nature having global
			information will aid in deblurring.
			
		\subsection{Scope}
			Deblurring is an active area of research in computer
			vision, new techniques and models are being developed
			in order to improve performance. Recent interest has
			been shown in unsupervised deblurring.
			
	\section{Literature Survey}
		\subsection{Information}
			Learning based methods used for Deblurring can be
			broadly classified into end-to-end and kernal estimation.
			In end-to-end given a blurred image, a sharp image is
			generated from it, it can be considered as a special case
			of image to image translation. In kernel estimation, the
			deep learning model is used to estimate the motion
			vectors in order to get the blur kernel, once the blur
			kernal is known the problem converts to non-blind
			deblurring and can be solved efficiently using classical
			methods. End to end methods tend to outperform kernel
			based methods.
			
			Most of the end to end methods have an encoder-
			decoder structure, where the encoder decreases the
			spatial dimension of the image while increasing its
			channels, in order to give a embedded representation of
			the image. This embedding is then used by the decoder
			to generate the sharp image, by increasing the spatial
			dimension of the image and reducing the number of
			channels, till they are equal to the original input.
			\cite{Noroozi}\cite{Nah_2017_CVPR}\cite{Tao_2018_CVPR} use scaled networks to improve
			performance, here they pass the image through the
			network each time increasing the resolution of input
			image, the scaled blurred image is concatenated with the
			restored image of the previous scale after it is up
			sampled.
			
			\cite{Kupyn_2018_CVPR}\cite{Ramakrishnan} uses a global skip, as blurred
			and deblurred are quite similar, it is better to learn how to
			restore only the blurred pixels, instead of the entire
			image. Therefore the blurred image is added to the last
			layer of the model, where the model only learns the
			corrections that is needed for each pixel. Almost all
			models used for Deblurring \cite{Kupyn_2018_CVPR}\cite{Nah_2017_CVPR}\cite{Tao_2018_CVPR} use
			local residual connections to avoid over fitting in deep
			networks.
			
			\cite{Ramakrishnan} in the generator of his model uses
			dense connections \cite{Huang_2017_CVPR}. Unlike CNN where
			input for a layer is the output of the previous layer, in
			dense connection input of a layer is the output of all the
			previous layers, concatenated and then squeezed. They
			enhance signal propagation and encourage feature
			reuse.
			
			\cite{Tao_2018_CVPR}\cite{Zhang_2018_CVPR} use recurrent networks to improve the
			long term dependencies of (global perception) of the
			models.
			
			It has been observed that the use of a classical loss
			function like L1, L2, or perceptual loss \cite{Johnson2016Perceptual},
			along with adversarial loss, improves the performance of
			GANs.
			
			Attention is used to improve the global perception of a
			model i.e the model learns which part of the image to
			give more attention to with respect to the others. There
			are two types attention used by us one is self-attention
			\cite{vaswani2017}\cite{sagan}\cite{nonlocal} and channel wise attention \cite{Hu2018SqueezeandExcitationN}. Details of these methods are given in Methodology Section.
			
			The evaluation of the performance regarding deblurring is
			done using metrics like PSNR and SSIM between the
			restored and the sharp (target) image.
		\subsection{Summary}
			Learining based methods can be classified into kernal
			estimation and end-to-end, among which end-to-end
			methods have shown better performance. Most of the
			models used in deblurring (or any image-image
			translation) have an encoder-decoder structure. Many
			techniques have been used in order to improve
			deblurring like use of scaled networks, skip connections,
			dense connections, recurrent modules and classical
			losses along with adversarial loss in GANs. Attention
			improves the global perception of a model and can be
			used to improve the performance of deblurring models.
			PSNR and SSIM are generally used to metrics to
			measure performance of a model.
	
	\section{Methodology}
		\subsection{Techniques to improve model performance}
			\subsubsection{Self Attention}
\begin{figure}[!h]
					\centering
					\includegraphics[width=0.8\linewidth]{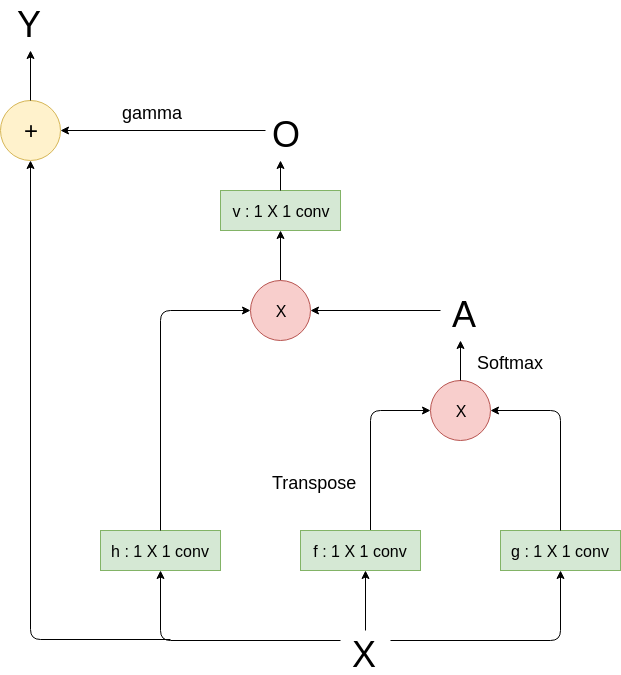}
					\caption{Representation of Self Attention, the symbols mean the same as in Eqn. \ref{eqn:self_attn}}
					\label{fig:self_attn}
				\end{figure}

				Self-Attention (Non-Local blocks) for images \cite{nonlocal} is similar to that of scaled dot product attention \cite{vaswani2017} used in language translation. Unlike
				CNN which captures information in a local neighborhood,
				non-local blocks capture long rage dependencies and
				give a global perception. It computes the attention at a
				position as the weighted sum of response at other
				positions.
				Given input features X, self attention can be shown as,
				\begin{equation}
					\begin{split}
						A = SoftMax(f(X)^T g(X))\\
						O = v(A h(X))\\
						Y = \gamma * O + X
					\end{split}
					\label{eqn:self_attn}
				\end{equation}
				
				where A denotes the attention map, and gamma is a
				scalar multiplied with O. All the functions f, g, h, v are
				implemented as 1 X 1 convolutions Fig. \ref{fig:self_attn}
			
			\subsubsection{Channel Attention}
			It finds the channels wise attention \cite{Hu2018SqueezeandExcitationN} of a feature.
			It consists of two components squeeze and excitation.
			Squeeze find the spatial average of each channel and
			gives an output Z according to
			\begin{equation}
				\begin{split}
					Z = GlobalAverage(X)\\
					where, Z_c = \frac{1}{H \times W} \sum_{i = 1}^{H} \sum_{j = 1}^{W} X_{i, j}
				\end{split}
				\label{eqn:squeeze}
			\end{equation}
			$Z_c$ denotes attention for each channel.
			After finding Z, we use the excitation block to find the
			attention of each channel i.e A given by
			\begin{equation}
				\begin{split}
					A = \sigma(W_2\delta(W_1 Z))\\
					Y = A * X
				\end{split}
				\label{eqn:excite}
			\end{equation}
			where $\sigma$ is sigmoid function and $\delta$ is the ReLU
			function, W1 and W2 denote the weights of two Fully
			Connected layers. We have implemented the fully
			connected layers as 1 X 1 convolution.
			\begin{figure}[!h]
				\centering
				\includegraphics[width=0.7\linewidth]{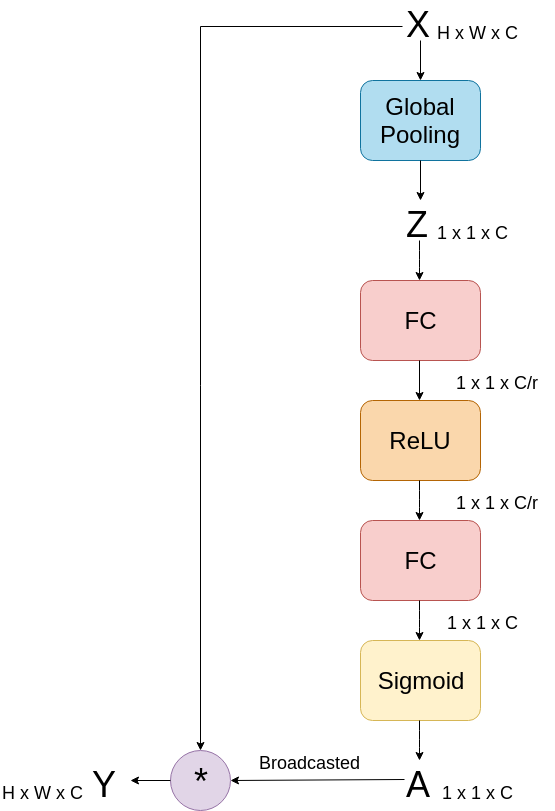}
				\caption{Representation of Channel Attention, the symbols mean the same as in Eqn. \ref{eqn:squeeze} Eqn. \ref{eqn:excite}. r is the factor by which the number of channels are initially decreased.}
				\label{fig:channel_attn}
			\end{figure}
		
		\subsubsection{Residual Learning (Skip Connections)}
			Residual learning can be divided into two types, Global
			and Local.
			
			In global residual learning we add the input image to the
			output of the model, it improves the performance of tasks
			like deblurring were the correlation between the blurred
			and sharp image is high. This requires the model to only
			learn the residuals, which are zero in many places, and
			increases the models learning capability.
			
			Local residual learning was introduced in ResNets. It
			increasing the learning capability of a model, by reducing
			over fitting, and allows us to make deeper models.
			
		\subsubsection{Spectral Normalization}
			It is a weight normalization technique \cite{miyato2018spectral} used to stabilize the training of GANs. The benefit of spectral normalization is that it only has one hyperparameter to tune i.e Lipschitz Constant and has a low computational cost. In the original paper \cite{miyato2018spectral} spectral normalization was only applied to the discriminator, but recent experiments in \cite{sagan} have shown that applying it to both generator and discriminator gives better results and hence we have done the same. 
			
		\subsubsection{Edge Information}
			To check weather edge information gives better
			performance, we extracted the edges of the sharp
			images using a Canny Edge Detector \cite{CANNY1986}, and concatenated
			it with the blurred image as an additional channel, and
			gave it as input to DeblurGAN model \cite{Kupyn_2018_CVPR}.
			
		\subsubsection{Feedback Module}
\begin{figure}[!h]
				\centering
				\includegraphics[width=\linewidth]{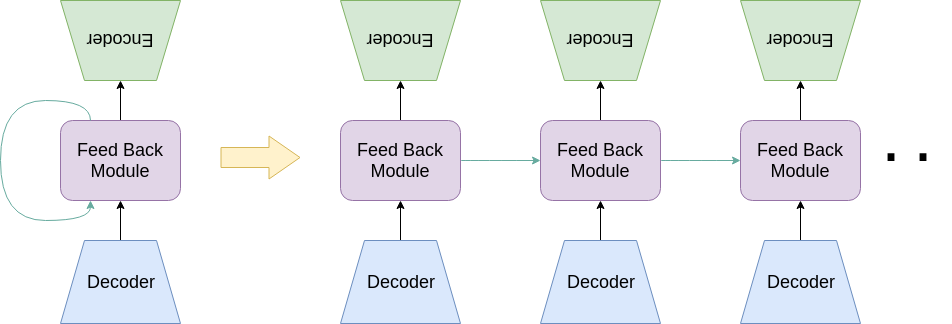}
				\caption{Representation of how a feedback network works.}
				\label{fig:feedback}
			\end{figure}

			LSTM is used between the encoder and decoder of a
			model to increase its long range dependency. The same image is iterated several times and each time the output of previous iteration's feedback module is given as input to the current feedback module as shown in Fig. \ref{fig:feedback}
		
		\subsubsection{Classical Losses}
			The classical losses we used with adversarial loss to help
			GAN training are L1, L2 and perceptual loss \cite{Johnson2016Perceptual}. L1 loss is the average of the absolute difference
			between the two images, it is determined using the below
			formula.
			\begin{equation}
				L_1 = \frac{\sum \lvert R - S \rvert}{H \times W \times C} 
			\end{equation}
			L2 loss is the average of the square of the difference
			between two images, as shown in the below.
			\begin{equation}
				L_2 = \frac{\sum (R - S)^2}{H \times W \times C} 
			\end{equation}
			It is similar to L2 loss expect that we use the features
			generated by a particular layer (like conv 3,3 ) of a
			particular model (like VGG19), instead of images, as
			shown below.
			\begin{equation}
				L_2 = \frac{\sum (\phi(R) - \phi(S))^2}{H \times W \times C} 
			\end{equation}
			In all the above formulas H, W, C denote the size of the
			dimensions. R, S denotes the restored (model output)
			and sharp (target) images. $\phi$ denotes the function (like
			conv 3, 3 of VGG19) that is used to generate the
			features.
			
	\subsection{Models used}
		\subsubsection{Pix2Pix}
			\begin{figure}[!h]
				\centering
				\includegraphics[width=\linewidth]{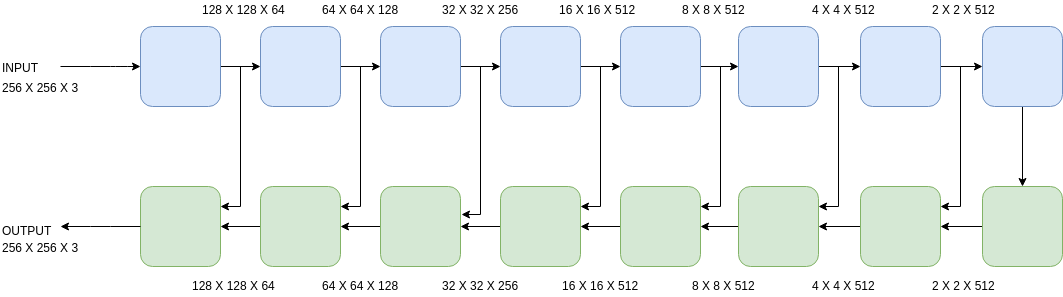}
				\caption{Representation of pix2pix network. Here the blue boxes denote the encoder blocks, while the green boxes denote the decoder blocks}
				\label{fig:pix2pix}
			\end{figure}
			The generator is an encoder decoder structure with skip
			connections from one encode block to a decode block,
			similar to U-Net \cite{unet} as shown in the figure Fig. \ref{fig:pix2pix}.
			Each encode block consists of a convolution with stride 2
			and padding “same” to reduce the spatial dimension by
			half. The number of channels are doubled until they they
			reach 512, except for the first encode block which
			
			increases the number of channels from 3 to 64. After the
			convolution we have a batch normalization and ReLU.
			Similarly each decode block consist of the transpose
			convolution, batch normalization and ReLU. Where the
			transpose convolution in contrast to convolution,
			increases the spatial dimension while decreasing the
			number of channels. The last decode block has a tanh
			activation instead of a ReLU.
			The discriminator used is a Markovian Discriminator
			(PatchGAN), it takes an N x N patch of a image and
			predicts weather it is the model output or the target, it
			does that for each patch and produces an image where
			each pixel denotes the prediction of the corresponding N
			x N patch. We average all the responses to get the final
			output.
			We used the techniques given above to improve this
			model for the task of deblurring, the exact details of which
			are in the Results section.
		\subsubsection{Residual in Residual Network}
			\begin{figure}[!h]
				\centering
				\begin{subfigure}{0.8\linewidth}
					\includegraphics[width=\linewidth]{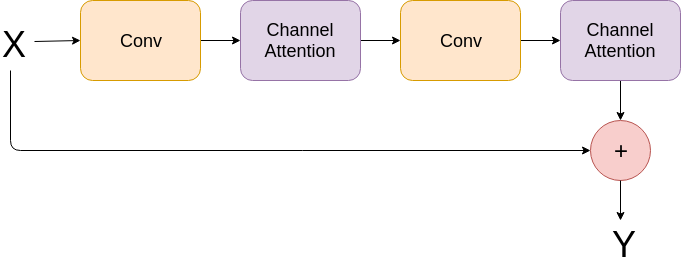}
					\caption{Residual Block \vspace{2em}}
					\label{fig:riraa}	
				\end{subfigure}
				
				\begin{subfigure}{0.48\linewidth}
					\includegraphics[width=\linewidth]{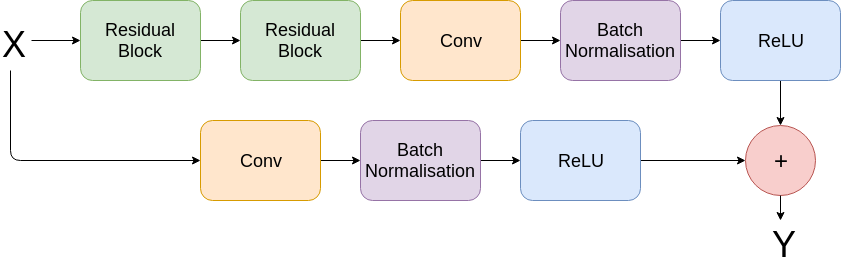}
					\caption{Encoder Module \vspace{2em}}
					\label{fig:rirab}
				\end{subfigure}
				\hfill
				\begin{subfigure}{0.48\linewidth}
					\includegraphics[width=\linewidth]{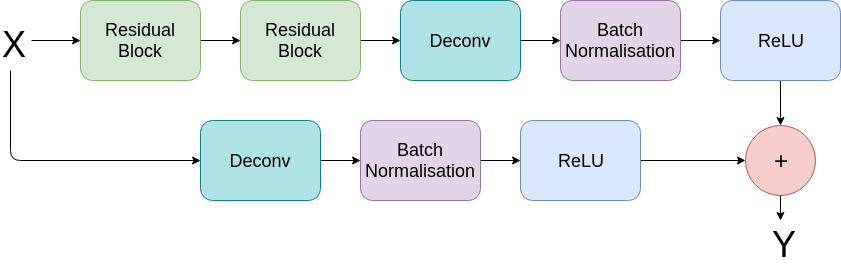}
					\caption{Decoder Module \vspace{2em}}
					\label{fig:rirac}
				\end{subfigure}
			
				\begin{subfigure}{\linewidth}
					\includegraphics[width=\linewidth]{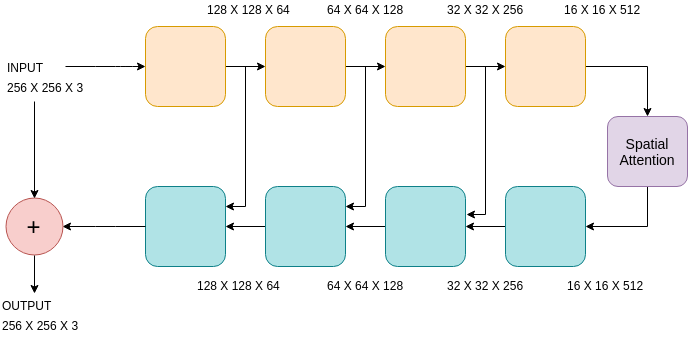}
					\caption{Residual in Residual model with channel and spatial attention, here the orange boxes denote encoder module, while the blue boxes denote the decoder module}
					\label{fig:rira}
				\end{subfigure}
			\end{figure}
			As the name suggests, several residual blocks are further
			given residual connections among them as shown in the
			Fig. \ref{fig:rirab} \ref{fig:rirac}. The networks design is inspired by \cite{Zhang_2018_ECCV}. The encoder decoder network with several
			encode and decode blocks is similar to that of pix2pix \cite{Isola_2017_CVPR}.
			But each encode/decode block is a residual in residual
			block as shown in Fig. \ref{fig:rira}. Each block consists of two Res
			blocks , a convolution/transpose convolution, a batch
			norm and a ReLU, the output is added with the input
			which has been scaled to the proper size after it passes
			through another convolution. Each of the Res Block
			consists of two series of convolution and channel
			
			attention, the output of which is added to the input.
			The model given above and a variant of it was
			implemented details of which are given in the Result
			section.
		
		\subsubsection{DeblurGAN}
			\begin{figure}[!h]
				\centering
				\includegraphics[width=\linewidth]{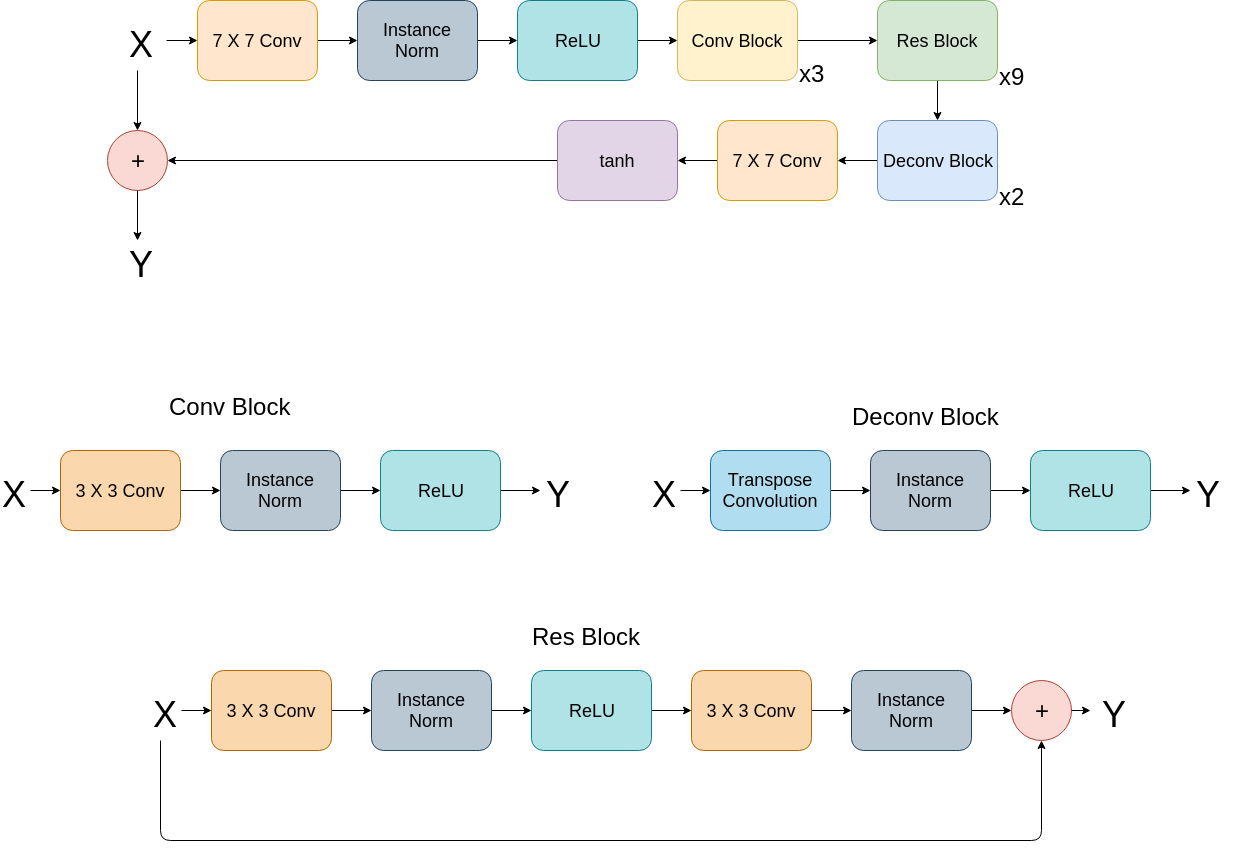}
				\caption{Representation of DeblurGAN \cite{Kupyn_2018_CVPR}}
				\label{fig:deblurnet}
				\end{figure}
			DeblurGAN \cite{Kupyn_2018_CVPR} structure is as shown in Fig.
			\ref{fig:deblurnet}. Techniques mentioned above were used to
			improve its performance, details of which are in the
			Result section.
	
	\subsection{Metrics}
		\subsubsection{PSNR}
			PSNR can be thought of as the reciprocal of MSE. MSE
			can be calculated as
			\begin{equation}
				MSE = \frac{\sum_{H, W}(S - R)^2}{H\times W}
			\end{equation}
			where H, W are the size of the dimensions of the image.
			S and R are the sharp and restored image respectively.
			Given MSE, PSNR can be calculated using,
			\begin{equation}
				PSNR = \dfrac{m^2}{MSE}
			\end{equation}
			where m is the maximum possible intensity value, since
			we are using 8-bit integer to represent a pixel in each
			channel, m = 255.
		\subsubsection{SSIM}
			SSIM helps us to find the structural similarity between
			two image, it can be calculated using,
			\begin{equation}
			SSIM(x, y) = \dfrac{(2\mu_x\mu_y + c_1)(2\sigma_{xy} + c_2)}{(\mu_x^2 + \mu_y^2 + c_1)(\sigma_x^2 + \sigma_y^2 + c_2)}
			\end{equation}
			where x, y are windows of equal dimension for restored and sharp images.
			$\mu_x, \mu_y$ denotes mean of x, y respectively. $\sigma_x, \sigma_y$ denotes variance for x, y respectively, whereas $\sigma_{xy}$ is the covariance between x and y. $c_1$ and $c_2$ are constants used to stabilize the division.
		
	\subsection{Dataset}
		GoPro \cite{Nah_2017_CVPR} was used for all experiments. To generate this dataset a GoPro Hero5 Black camera was used which captured high resolution (1280 × 720), high frame rate (240 frames per second) videos outdoor. 
		To generate blurred image an average of a few frames (odd number picked randomly from 7 to 23) was taken, while the central frame was considered as the corresponding sharp image. To reduce the magnitude of relative motion across frames they were down sampled and to avoid artifacts caused by averaging only frames were the optical flow was at most 1 were considered.
\section{Results}
	\subsection{Pix2pix}
		\begin{table}[h]
			\begin{center}
				\label{table:pix2pix}
				\begin{tabular}{|c | c | c |} 
					\hline
					\textbf{Methods}& \textbf{PSNR} & \textbf{SSIM} \\  
					\hline\hline
					\textit{Pix2Pix} & 25.41 & 0.810  \\ 
					\hline
					\textit{Pix2Pix + Sl. Attn} & 25.41 & 0.808  \\
					\hline
					\textit{Pix2Pix + Sl. Attn + GR} & \textbf{27.29} & \textbf{0.858} \\
					\hline
					\textit{Pix2Pix + Sl. Attn + SN} & 26.05 & 0.831  \\
					\hline
					\textit{Pix2Pix + Sl. Attn + Ch. Attn + GR + SN + Percep. Loss} & 27.05 & 0.831  \\  
					\hline
				\end{tabular}
				\caption{PSNR and SSIM for different improvements on the pix2pix model. Sl. Attn refers to Self Attention. GR refers to Global Residual. SN refers to spectral normalization. Ch. Attn refers to Channel. Percep. Loss refers to perceptual loss}
			\end{center}
		\end{table}
	In case of addition of attention (only spatial or both spatial and channel), they were added after
	every three encoder/decoder blocks, with a channel attention between the encoder and the
	decoder when the feature size 1 x 1 x 512. All test were done on 256 X 256 images. Each model was trained for 50 epochs.
	\subsection{Residual in Residual}
		\begin{table}[h]
			\begin{center}
				\label{table:rira}
				\begin{tabular}{|c | c | c |} 
					\hline
					\textbf{Methods}& \textbf{PSNR} & \textbf{SSIM} \\  
					\hline\hline
					\textit{RiR + Ch. Attn + Sl. Attn + Percep. Loss + SN + GR} & 22.70 & 0.640  \\ 
					\hline
					\textit{RiR(Large) + Ch. Attn + Sl. Attn + L1 Loss + SN + GR} & \textbf{23.46} & \textbf{0.671}  \\
					\hline
				\end{tabular}
				\caption{PSNR and SSIM for different improvements on the Residual in Residual (RiR) model. Sl. Attn refers to Self Attention. GR refers to Global Residual. SN refers to spectral normalization. Ch. Attn refers to Channel. Precep. Loss refers to perceptual loss}
			\end{center}
		\end{table}
		The “large” variant is similar to Fig. \ref{fig:rira}, except that it uses one extra encoder as well as decoder
		block. All models are trained for 300 epochs. RiR model is tested on 1280 X 720 images, while RiR(Large) is tested on 1280 X 768 images.
	\subsection{DeblurGAN}
			\begin{table}[h]
				\begin{center}
					\label{table:deblurgan}
					\begin{tabular}{|c | c | c |} 
						\hline
						\textbf{Methods}& \textbf{PSNR} & \textbf{SSIM} \\  
						\hline\hline
						\textit{DeblurGAN} & \textbf{28.70} & \textbf{0.958}  \\ 
						\hline
						\textit{DeblurGAN + Edge Information} & 25.27 & 0.773  \\
						\hline
						\textit{DeblurGAN + Feedback} & 27.20 & 0.827  \\
						\hline
					\end{tabular}
					\caption{PSNR and SSIM for different improvements on the DeblurGAN model}
				\end{center}
			\end{table}
			Each image is iterated 4 times over the feedback module. Tests are done on 1280 X 720 images. Each model is trained for 300 epochs.
			
\section{Conclusion}
	Different GAN models were used for the task of deblurring and there performance was improved using various techniques mentioned above. Residual connections (specially global residual) and attention modules, have shown an improvement in performance to the existing model. Use of classical losses and spectral normalization were also helpful for stable GAN training. Use of larger models gave better performance (like in case of RiR and RiR(Large)). Use of edge information and feedback modules seems not to improve the performance of the model. The implementation of the models can be found in the given link  \href{https://github.com/lenka98/Bind-Deblurring-using-GANs}{https://github.com/lenka98/Bind-Deblurring-using-GANs}
\section*{Acknowledgment}
I would like to thank Prof. Anurag Mital and Anubha Pandey for their guidance and support. I
would also like to thank IAS and IIT Madras for giving me this opportunity.

\bibliographystyle{plain}
\bibliography{refs}

\newpage
\section*{Appendices}
\appendix
\section{Abbreviations}
Below are the abbreviations used:
\begin{table}[h]
	\begin{center}
		\label{table:abbre}
		\begin{tabular}{|c | c |} 
			\hline
			\textit{GAN}& Generative Adversarial Network \\  
			\hline
			\textit{CNN} & Convolutional Neural Network \\ 
			\hline
			\textit{PSNR} & Peak Signal to Noise Ratio \\
			\hline
			\textit{SSIM} & Structural Similarity \\
			\hline
			\textit{ReLU} & Rectified Linear Unit \\
			\hline
			\textit{FC} & Fully Connected \\
			\hline
			\textit{LSTM} & Long Short Term Memory \\
			\hline
			\textit{MSE} & Mean Square Error \\
			\hline
		\end{tabular}
	\end{center}
\end{table}

\section{Notations}
Below are the Notations used.
\begin{table}[h]
	\begin{center}
		\label{table:abbre}
		\begin{tabular}{|c | c |} 
			\hline
			\textit{B}& Blurred Image \\  
			\hline
			\textit{S} & Sharp Image (Ground Truth) \\ 
			\hline
			\textit{R} & Restored Image (Model Output) \\
			\hline
			\textit{K} & Blur Kernel \\
			\hline
			\textit{N} & Additive Noise\\
			\hline
			\textit{FC} & Fully Connected \\
			\hline
			\textit{X} & Input to a module \\
			\hline
			\textit{Y} & Output of a module \\
			\hline
			\textit{A} & Attention Map \\
			\hline
			\textit{O} & Intermediate output \\
			\hline
			\textit{$\gamma$} & Learnable constant \\
			\hline
			\textit{Z} & Squeezed Features \\
			\hline
			\textit{$\sigma$} & Sigmoid Activation \\
			\hline
			\textit{$\delta$} & ReLU Activation \\
			\hline
			\textit{$W_i$} & Weight Matrix \\
			\hline
			\textit{H, W, C} & Height, Width, and Channels of a feature/image \\
			\hline
			\textit{$\phi$} & Function representing a CNN layer \\
			\hline
			\textit{$\mu_x$} & Mean of x \\
			\hline
			\textit{$\sigma_x$} & Variance of x \\
			\hline
			\textit{$\sigma_{xy}$} & Co-Variance of x w.r.t y \\
			\hline
		\end{tabular}
	\end{center}
\end{table}

\end{document}